# Diversity-Integration Trade-offs in MIMO Detection

Antonio De Maio*, Marco Lops†, and Luca Venturino†
*Università degli Studi di Napoli "Federico II", Napoli, ITALY – E-mail: ademaio@unina.it
†Università degli Studi di Cassino, Cassino (FR), ITALY – E-mail: lops@unicas.it, l.venturino@unicas.it

*Abstract*— In this work, a MIMO detection problem is considered. At first, we derive the Generalized Likelihood Ratio Test (GLRT) for arbitrary transmitted signals and arbitrary time-correlation of the disturbance. Then, we investigate design criteria for the transmitted waveforms in both power-unlimited and power-limited systems and we study the interplay among the rank of the optimized code matrix, the number of transmit diversity paths and the amount of energy integrated along each path. The results show that increasing the rank of the code matrix allows generating a larger number of diversity paths at the price of reducing the average signal-to-clutter level along each path.

## I. INTRODUCTION

We consider a detection system wherein both the receiver and the transmitter, which are not necessarily co-located, are equipped with multiple widely spaced antennas. The scenario is that typical of a Multiple-Input Multiple-Output (MIMO) radar architecture, as outlined in [1] and analyzed in [2], [3], but may also describe the tasks of a sensor network with widely spaced nodes, wherein both the transmit and the receive array elements exhibit no coherence. In this situation, the MIMO architecture can be advocated to improve the detectability of targets with fluctuating radar cross section either through an increase of the diversity order [1], [2] or through beam-forming in the signal space [3] or through a combination of these two strategies: the degree of freedom which allows jumping from one strategy to another is the structure of the transmitted signals, and in particular the rank of the space-time code matrix used at the transmitter.

In the above context, this paper makes the following contributions. At the receiver-design level, we first derive the Generalized Likelihood Ratio Test (GLRT) for arbitrary transmitted waveforms and arbitrary time-correlation of the disturbance. Building upon [2], [3], two design criteria for the space-time code are then presented and discussed. The figure of merit is the Mutual Information (MI) [4] between the received signals and the channel vectors generated by a prospective target, whose maximization is related to the ability to estimate unknown characteristics of the target as shown in [5], and also to the maximization of the detection probability ($P_d$) for a fixed false alarm probability ($P_{fa}$) as pointed out in [2], [3]. We examine both the case of power-unlimited systems, wherein a constraint is forced upon the *received* average signal-to-disturbance ratio, and the case of power-limited systems, wherein the constraint is instead on the *transmitted* energy. Finally, closed-form formulas for $P_{fa}$ and $P_d$ are given, showing that there is an inherent trade-off between number of diversity paths and amount of energy integration along each path, which is a direct confirmation of the well-known fact that no uniformly optimum (e.g., for any signal-to-disturbance ratio) coding strategy exists. The tool that can be used for trading diversity order for integration is the code-matrix structure, and particularly its rank.

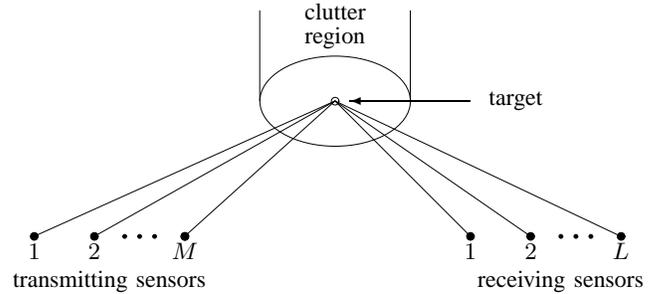

Fig. 1. Considered scenario.

This paper is organized as follows. In Section II, the problem under investigation is described. In Section III, the GLRT-based detector is derived. Finally, the diversity-integration trade-offs for different code design criteria are analyzed in Section IV.

## II. PROBLEM STATEMENT

Consider a detection architecture where the transmitter and the receiver consist of $M$ and $L$ widely spaced antennas, respectively, and are not necessarily co-located as depicted in Figure 1. We assume the following general model for the signal transmitted by the $m$-th antenna (node)

$$s_m(t) = \sum_{n=1}^{N} a_{n,m} \phi_n(t), \quad 0 \leq t \leq T_s, \ m=1,\ldots,M, \quad (1)$$

where $N$ is the signal-space dimension, $\{\phi_n(t)\}_{n=1}^{N}$ is an orthonormal basis of the considered signal space, $\boldsymbol{a}_m = (a_{1,m},\ldots,a_{N,m})^T$ is the codeword associated with the $m-$th transmitter, $\|\boldsymbol{a}_m\|^2$ is the energy transmitted by antenna $m$ and $T_s$ is the time duration of the transmitted signal. Subsumed by the above model is the case of a pulsed-MIMO radar, wherein $\phi_n(t) = p(t-(n-1)T)$, with $p(\cdot)$ a pulse of duration $T_p \leq T$ and $T = T_s/N$ the pulse-repetition time (PRT) [2].

The signal observed at the $L$ receive nodes may contain or not the echo of a target scattering located at a given distance. The bandwidth of the transmitted signal induces a partition of the controlled area in a finite number of range cells, directly tied to the delay with which a target echo is heard at the receiving nodes. Denoting by $r_\ell(t)$ the signal received at the $\ell-$th receive node at a given delay $\tau_0$, we thus have:

$$r_\ell(t) = \begin{cases} H_1 : \sum_{m=1}^{M}\sum_{n=1}^{N} a_{n,m}\phi_n(t-\tau_0)\alpha_{m,\ell} + w_\ell(t) \\ H_0 : w_\ell(t) \end{cases} \quad (2)$$

for $\ell = 1, \ldots, L$. In the above equation, which subsumes the model outlined in [2], $w_\ell(t)$ is a Gaussian process which models the overall disturbance originating from the superposition of receiver noise and spurious scattering from the surrounding environment (denoted as clutter) which typically exhibits time correlation. In writing down (2), the following three assumptions have been made. a) The transmitted signals are narrowband [1], namely all of the receive sensors see the target as belonging to the same range cell[1]. b) The spacing of the antennas at both the transmitter and the receiver is wide enough as to allow *angle diversity* [1], [6], i.e., to generate different aspect angles which justifies the use of the $LM$ different coupling coefficients $\alpha_{m,\ell}$ for $m = 1, \ldots, M$ and $\ell = 1, \ldots, L$. c) The target is either stationary or has a known Doppler shift which is compensated for at the receiver [7].

In principle, Analog-to-Digital (A/D) conversion would require undertaking a Kelly-Root expansion of the received (vector) signal [8], which would inevitably lead to an A/D stage depending on the clutter covariance[2]. This would in turn lead to a hardly implementable receiver and would prevent any subsequent step towards adaptive systems, wherein such a covariance is assumed unknown [3]. As a consequence, we adopt the customary - albeit sub-optimum - approach of projecting the received observations onto the signal space, which yields the following binary hypothesis testing problem:

$$r_\ell = \begin{cases} H_1 : A\alpha_\ell + w_\ell \\ H_0 : w_\ell \end{cases} \quad \ell = 1, \ldots, L, \quad (3)$$

where $\alpha_\ell = (\alpha_{1,\ell}, \ldots, \alpha_{M,\ell})^T$ is the $M$-dimensional vector modeling the scattering of the target - hit by the $M$ transmitted waveforms in (1) - towards the $\ell$-th receive node for $\ell = 1, \ldots, L$; $A = (a_1, \ldots, a_M)$ is the $N \times M$ *code matrix* containing the $N$-dimensional codeword transmitted by the $m$-th transmit node as its $m$-th column; $\text{tr}\{AA^H\}$ is the total transmitted energy; finally, $w_\ell$ represents the overall disturbance at the $\ell$-th receive antenna and is modeled as an $N$-dimensional Gaussian vector with known (full-rank) correlation matrix $M$.

For future reference, we define $\Delta = \min\{N, M\}$ and $\delta = \text{rank}\{A\}$, with $1 \leq \delta \leq \Delta$. Also, we define the received average signal-to-clutter ratio (SCR) as

$$\overline{\text{SCR}} = \frac{1}{NL} \mathbb{E}_\alpha \left[ \sum_{\ell=1}^L \alpha_\ell^H A^H M^{-1} A \alpha_\ell \right] = \frac{\sigma_\alpha^2}{N} \text{tr}\{A^H M^{-1} A\}, \quad (4)$$

where the expectation is over the realizations of the vector $\alpha = \text{vec}\{(\alpha_1, \ldots, \alpha_L)\}$, $\text{tr}\{\cdot\}$ denotes trace and the assumption has been made that $\mathbb{E}[\alpha_\ell \alpha_\ell^H] = \sigma_\alpha^2 I_M$ for $\ell = 1, \ldots, L$.

Given the above ingredients, we first derive a distribution-free test for (3), namely a test whose implementation does not require prior knowledge of the target parameters: in this work, we assume that detection takes place at a fusion center wherein un-quantized versions of the $r_\ell$'s are made available. Next, we investigate two design criteria for the code matrix $A$, so as to elicit the interplay between the transmitted waveforms $s_1(t), \ldots, s_M(t)$ and detection performance.

---
[1]This assumption can be relaxed through cell synchronization.
[2]This becomes the conventional Karhunen-Loewe expansion to be undertaken at each receive antenna when, as assumed throughout this study, the clutter exhibits temporal correlation, but no spatial correlation.

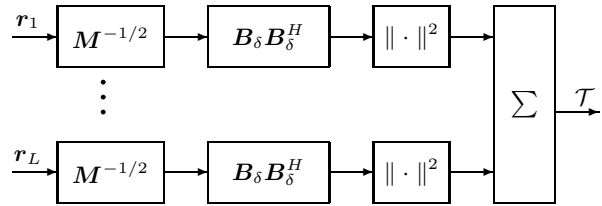

Fig. 2. GLRT statistic computation.

### III. DETECTOR STRUCTURE

Lacking prior information as to the target statistics, the Neyman-Pearson test cannot be implemented, whereby a wise design strategy for the detector is the Generalized Likelihood Ratio Test (GLRT) [7] which, under the model (3), yields:

$$\mathcal{T} = \max_{\alpha_1, \ldots, \alpha_L} \ln \left[ \frac{f(r_1, \ldots, r_L | H_1, \alpha_1, \ldots, \alpha_L)}{f(r_1, \ldots, r_L | H_0)} \right] \underset{H_0}{\overset{H_1}{\gtrless}} \eta, \quad (5)$$

where $f(r_1, \ldots, r_L | H_1, \alpha_1, \ldots, \alpha_L)$ and $f(r_1, \ldots, r_L | H_0)$ denote the conditional densities of the observations under the two alternative hypotheses, while $\eta$ is the detection threshold, to be set based upon the desired false alarm probability, $P_{fa}$ say. The maximization involved by (5) is obviously tied to the structure of $A$, which can be in principle arbitrary, and in particular to its rank $\delta$, which in turn may take on any integer value from 1 to $\Delta$. Straightforward derivations lead to the following general form of the GLRT, which holds for *arbitrary* code-matrix $A$:

$$\mathcal{T} = \sum_{\ell=1}^L \left\| P_{M^{-1/2}A} M^{-1/2} r_\ell \right\|^2 \underset{H_0}{\overset{H_1}{\gtrless}} \eta, \quad (6)$$

wherein $P_{M^{-1/2}A}$ represents the orthogonal projector onto the range span of the matrix $M^{-1/2}A$.

Let $BDC^H$ be the singular value decomposition (SVD) [9] of $M^{-1/2}A$, where $B$ is a unitary $N \times N$ matrix containing the left singular vectors, $D$ is an $N \times M$ matrix containing the singular values on the main diagonal with $[D]_{1,1} \geq \cdots \geq [D]_{\Delta,\Delta} \geq 0$, and $C$ is a unitary $M \times M$ matrix containing the right singular vectors. Under rank-$\delta$ coding, the orthogonal projector in (6) can be written as $P_{M^{-1/2}A} = B_\delta B_\delta^H$, wherein $B_\delta$ is the $N \times \delta$ matrix containing the first $\delta$ columns of $B$. Three relevant cases are subsumed by this solution.

1) Under rank-1 coding, we have that $A = aq^H$ with $a$ $N$-dimensional and $q$ unit-energy $M$-dimensional; hence, in this case the orthogonal projector reduces to

$$P_{M^{-1/2}A} = \left( a^H M^{-1} a \right)^{-1} M^{-1/2} aa^H M^{-1/2}.$$

2) Under rank-$\Delta$ (i.e., full-rank) coding and $N \leq M$, we have $P_{M^{-1/2}A} = I_N$.

3) Finally, under rank-$\Delta$ coding and $N \geq M$, we have

$$\boldsymbol{P}_{\boldsymbol{M}^{-1/2}\boldsymbol{A}} = \boldsymbol{M}^{-1/2}\boldsymbol{A}(\boldsymbol{A}^H\boldsymbol{M}^{-1}\boldsymbol{A})^{-1}\boldsymbol{A}^H\boldsymbol{M}^{-1/2}.$$

Notice that $\boldsymbol{M}^{-1/2}\boldsymbol{r}_\ell$ represents the output of a filter aimed at whitening the disturbance impinging on the $\ell$-th transmit antenna, whereby the computation of the GLRT test statistic in (6) may be undertaken based on the scheme of Figure 2.

The performances of the above test can be expressed in closed - albeit implicit - form under very general conditions. In particular, under rank-$\delta$ coding, we have:

$$P_{fa} = e^{-\eta} \sum_{k=0}^{\delta L - 1} \frac{\eta^k}{k!}, \tag{7}$$

$$P_d = \mathrm{E}_{\boldsymbol{\alpha}}\left[Q_{\delta L}\left(\sqrt{2\beta},\sqrt{2\eta}\right)\right], \tag{8}$$

where $\beta = \sum_{\ell=1}^L \boldsymbol{\alpha}_\ell^H \boldsymbol{A}^H \boldsymbol{M}^{-1} \boldsymbol{A} \boldsymbol{\alpha}_\ell$ and $Q_m(\cdot,\cdot)$ denotes the generalized Marcum Q function of order $m$. Notice that (7) and (8) generalize previous results derived in [2] for $\delta = \Delta$ and in [3] for $\delta = 1$, respectively.

IV. CODE DESIGN AND PERFORMANCE BOUNDS

Code optimization requires determining a meaningful performance measure to be optimized. Paralleling the arguments of [2], we assign a prior to the vectors of the scattering coefficients; in particular, we assume that $\boldsymbol{\alpha}$ consists of $LM$ independent and identically distributed circularly-symmetric Gaussian random variables with equal variance $\sigma_\alpha^2$. At this point, a reasonable design criterion in the GLRT framework is to choose $\boldsymbol{A}$ so as to maximize the MI between the received observations $\boldsymbol{r}_1,\ldots,\boldsymbol{r}_L$ and the channel vectors $\boldsymbol{\alpha}_1,\ldots,\boldsymbol{\alpha}_L$ under $H_1$, i.e., $I(\boldsymbol{r}_1,\ldots,\boldsymbol{r}_L;\boldsymbol{\alpha}_1,\ldots,\boldsymbol{\alpha}_L|H_1) = L\log_2\left[\det\left(\boldsymbol{I}_N + \sigma_\alpha^2 \boldsymbol{M}^{-\frac{1}{2}}\boldsymbol{A}\boldsymbol{A}^H\boldsymbol{M}^{-\frac{1}{2}}\right)\right]$ [4]. Indeed, since the GLRT amounts to writing the conditional likelihood given the unknown target parameters and replacing them through their maximum-likelihood estimates performed based upon the available observations, maximizing the MI between the received signals and the channel vectors is expected to improve the GLRT performance [5].

To make the maximization non-trivial, a set of physical constraints on $\boldsymbol{A}$ should be added. In the following we discuss two relevant cases: I) a power-unlimited system, wherein we constrain the normalized received average SCR in (4); II) a power-limited system, wherein we constrain the transmitted energy per signal dimension, i.e., $\mathrm{tr}\{\boldsymbol{A}\boldsymbol{A}^H\}/N$. In both cases, an additional constraint can be imposed upon the rank of $\boldsymbol{A}$.

A. Power-unlimited system

The problem to be solved is the following:

$$\begin{cases} \max_{\boldsymbol{A}} \log_2\left[\det\left(\boldsymbol{I}_N + \sigma_\alpha^2 \boldsymbol{M}^{-\frac{1}{2}}\boldsymbol{A}\boldsymbol{A}^H\boldsymbol{M}^{-\frac{1}{2}}\right)\right] \\ \text{s.t. } \overline{\mathrm{SCR}} = \frac{\sigma_\alpha^2}{N}\mathrm{Tr}\{\boldsymbol{A}^H\boldsymbol{M}^{-1}\boldsymbol{A}\} \leq \nu \\ \mathrm{rank}\{\boldsymbol{A}\} \leq \theta \end{cases} \tag{9}$$

with $\nu > 0$ and $\theta \in \{1, 2, \ldots, \Delta\}$. Let $\boldsymbol{B}\boldsymbol{\Lambda}\boldsymbol{B}^H$ be the spectral decomposition of the positive semi-definite matrix $\boldsymbol{T} = \boldsymbol{M}^{-\frac{1}{2}}\boldsymbol{A}\boldsymbol{A}^H\boldsymbol{M}^{-\frac{1}{2}}$, where $\boldsymbol{\Lambda} = \boldsymbol{D}\boldsymbol{D}^H = \mathrm{diag}\{\lambda_1, \ldots, \lambda_N\}$ with $\lambda_1 \geq \ldots \geq \lambda_N \geq 0$. Since the rank constraint forces $\lambda_{\theta+1} = \ldots = \lambda_N = 0$, the problem (9) can also be recast as

$$\begin{cases} \max_{\lambda_1 \geq 0, \ldots, \lambda_\theta \geq 0} \sum_{j=1}^{\theta} \log_2(1 + \sigma_\alpha^2 \lambda_j) \\ \text{s.t. } \overline{\mathrm{SCR}} = \frac{\sigma_\alpha^2}{N}\sum_{j=1}^{\theta} \lambda_j \leq \nu \end{cases} \tag{10}$$

A straightforward application of the Jensen inequality [4] allows now deriving the following condition for optimality:

$$\lambda_j = \frac{N\nu}{\theta\sigma_\alpha^2} = \frac{N\overline{\mathrm{SCR}}}{\theta\sigma_\alpha^2}, \quad j = 1,\ldots,\theta, \tag{11}$$

which in turn implies that the optimal $\boldsymbol{A}$ should be such that

$$\boldsymbol{M}^{-\frac{1}{2}}\boldsymbol{A}\boldsymbol{A}^H\boldsymbol{M}^{-\frac{1}{2}} = \frac{N\overline{\mathrm{SCR}}}{\theta\sigma_\alpha^2}\boldsymbol{B}_\theta\boldsymbol{B}_\theta^H. \tag{12}$$

According to (11) and (12), the optimal space-time code must have $\mathrm{rank}\{\boldsymbol{A}\} = \delta = \theta$ (i.e., the code must exploit the maximum possible number of degrees of freedom as permitted by the rank constraint) and $\overline{\mathrm{SCR}} = \nu$; whereby, both inequality constraints in (9) can equivalently be replaced with a strict equality. The following remarks are now in order.

For a target $\overline{\mathrm{SCR}}$ and a given rank constraint $\theta$, the above coding strategy amounts to generating $\theta$ independent and identically distributed diversity paths at each receive antenna; also, each path enjoys the same average signal-to-clutter ratio given by $N\overline{\mathrm{SCR}}/\theta$. Notice that increasing the number of diversity paths $\theta$ at each receive antenna (i.e., increasing the rank of $\boldsymbol{A}$) comes at the price of reducing the received average SCR per path. This intuition will be farther exploited in what follows.

The solution in (11) coincides with the solution that we would obtain if, under a definite rank constraint, we adopted the lower Chernoff-bound on the detection probability as objective function. This result was first observed in [2] for full-rank coding (i.e., $\theta = \Delta$).

Finally, we point out that the optimal code matrix complying with (12) is not unique. In fact, let $\boldsymbol{U}_A\boldsymbol{\Sigma}_A\boldsymbol{V}_A^H$ be the SVD of $\boldsymbol{A}$, where $\boldsymbol{U}_A$ is a unitary $N \times N$ matrix containing the left singular vectors, $\boldsymbol{\Sigma}_A$ is an $N \times M$ matrix containing the singular values on the main diagonal and $\boldsymbol{V}_A$ is a unitary $M \times M$ matrix containing the right singular vectors. Also, let $\boldsymbol{U}_M\boldsymbol{\Lambda}_M^{-1}\boldsymbol{U}_M^H$ be the spectral decomposition of $\boldsymbol{M}^{-1}$, where $\boldsymbol{U}_M = (\boldsymbol{u}_{M,1},\ldots,\boldsymbol{u}_{M,N})$ is an $N \times N$ unitary matrix and $\boldsymbol{\Lambda}_M^{-1} = \mathrm{diag}\{1/\lambda_{M,1},\ldots,1/\lambda_{M,N}\}$ with $1/\lambda_{M,1} \geq \ldots \geq 1/\lambda_{M,N} > 0$. Then, (12) can be rewritten as

$$\boldsymbol{U}_M\boldsymbol{\Lambda}_M^{-1/2}\boldsymbol{U}_M^H\boldsymbol{U}_A\boldsymbol{\Sigma}_A = \sqrt{\frac{N\overline{\mathrm{SCR}}}{\theta\sigma_\alpha^2}}\boldsymbol{B}_\theta. \tag{13}$$

It is clear that the unitary matrix $\boldsymbol{V}_A$ can be arbitrarily chosen since it does not come into play in (13). Also, it is instructive to consider the class of optimal solutions obtained by setting $\boldsymbol{U}_A = \boldsymbol{U}_M$. In this case, the right singular vectors of $\boldsymbol{A}$ are matched to the $N$ eigenvectors of the noise matrix $\boldsymbol{M}$ which in turn define as many orthogonal modes in the signal space. The position of the $\theta$ non-zero singular values in $\boldsymbol{\Sigma}_A$ determines which subset of orthogonal modes is employed for transmission. In particular, let $n(1),\ldots,n(\theta)$ be the row (or column) indexes corresponding to the non-zero singular values in $\boldsymbol{\Sigma}_A$. Condition (13) is fulfilled by setting

$$[\boldsymbol{\Sigma}_A]_{n(j),n(j)} = \sqrt{\lambda_{M,n(j)} \frac{N\overline{\text{SCR}}}{\theta \sigma_\alpha^2}}, \quad j=1,\ldots,\theta. \quad (14)$$

It is seen from (14) that more energy must be allocated to more interfered modes in order to equalize their received average SCR's. Among all the possible choices, setting $n(j) = j$ for $j=1,\ldots,\theta$ minimizes the total transmit energy $\text{tr}\{\boldsymbol{\Sigma}_A \boldsymbol{\Sigma}_A^H\}$ required to achieve a given $\overline{\text{SCR}}$ since transmission takes place along the least interfered modes.

*1) Analysis and trade-offs:* The detection performance of the GLRT under rank-$\theta$ optimal coding in (11) has a simple closed-form expression. Indeed, since the constraint on $\overline{\text{SCR}}$ allows generating exactly $\theta L$ independent diversity paths with one and the same received average SCR given by $N\overline{\text{SCR}}/\theta$, the test statistic (6) is a Gamma random variable with shape parameter $\theta L$ and scale parameter $1 + N\overline{\text{SCR}}/\theta$ under $H_1$, implying that

$$P_d = \exp\left\{-\frac{\eta}{1+N\overline{\text{SCR}}/\theta}\right\} \sum_{\ell=0}^{\theta L - 1} \frac{1}{\ell!} \left(\frac{\eta}{1+N\overline{\text{SCR}}/\theta}\right)^\ell.$$

The inherent trade-off between number of generated diversity paths $\theta L$ and amount of energy integration granted on each path $N\overline{\text{SCR}}/\theta$ is visible already at this stage, and is confirmed by the asymptotic behavior (for increasingly large $\overline{\text{SCR}}$) of the detection probability, i.e.:

$$P_d \sim 1 - \frac{1}{(\theta L)!}\left(\frac{\eta}{1+N\overline{\text{SCR}}/\theta}\right)^{\theta L}. \quad (15)$$

Notice that no strategy is uniformly superior in terms of detection performance, but the optimal value of $\theta$ depends upon the operating $\overline{\text{SCR}}$, as also confirmed by Figure 3. Indeed, while for large $\overline{\text{SCR}}$ the asymptotic behavior in (15) indicates that maximizing the diversity order (i.e., choosing $\theta = \Delta$) amounts to maximizing the detection probability, in the low signal-to-clutter regime this trend does not hold anymore. To have an intuitive justification of this fact, recall that increasing the number of diversity paths eventually leads to more and more constrained target amplitude fluctuations. On the other hand, (8) reveals that the detection probability is in the form $P_d = \text{E}_\beta[F(\beta,\eta)] = \text{E}_{\beta_0}[F(\beta_{rms}\beta_0,\eta)]$, where $\beta_{rms}$ is the root mean square value of the random variable $\beta$, while $\beta_0$ is a random variable with unit rms value. $F(\beta,\eta)$, regarded as a function of $\beta$, is a cumulative distribution function and, therefore, exhibits a sigmoidal shape (i.e., is $\cup$-convex in the region $\beta_{rms} \to 0$, while being $\cap$-convex in the region $\beta_{rms} \to \infty$). Under these circumstances, more and more constrained fluctuations (i.e., lower dispersion ratios of the parameter $\beta$) typically result in larger values of $P_d$ in the $\cap$-convexity region, while being detrimental in the $\cup$-convexity region.

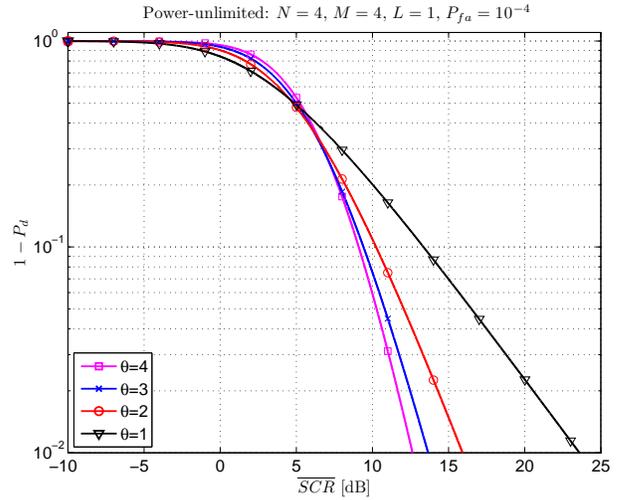

Fig. 3. Probability of miss versus $\overline{\text{SCR}}$ for the optimal coding in (11).

### B. Power-limited system

In power-limited systems, the problem to be solved is

$$\begin{cases} \max_{\boldsymbol{A}} \quad \log_2\left[\det\left(\boldsymbol{I}_N + \sigma_\alpha^2 \boldsymbol{M}^{-\frac{1}{2}}\boldsymbol{A}\boldsymbol{A}^H \boldsymbol{M}^{-\frac{1}{2}}\right)\right] \\ \text{s.t.} \quad \text{tr}\{\boldsymbol{A}\boldsymbol{A}^H\} \leq N\mathcal{E} \\ \qquad \text{rank}\{\boldsymbol{A}\} \leq \theta \end{cases} \quad (16)$$

where $\mathcal{E} > 0$ is the available transmit energy per signal dimension and $\theta \in \{1,2,\ldots,\Delta\}$. A solution to (16) was derived in [3] for $\theta = N \leq M$. Also, maximizing the objective function in (16) for $\theta = 1$ amounts to maximizing $\overline{\text{SCR}}$ in (4) and the solution to this problem under a transmit energy constraint can again be found in [3]. In the following, the solution to (16) for arbitrary values of $\theta$ is discussed.

Notice first that the problem (16) can be recast as

$$\begin{cases} \max_{\boldsymbol{T} \succeq 0} \quad \log_2\left[\det\left(\boldsymbol{I}_N + \sigma_\alpha^2 \boldsymbol{T}\right)\right] \\ \text{s.t.} \quad \text{tr}\{\boldsymbol{M}^{\frac{1}{2}}\boldsymbol{T}\boldsymbol{M}^{\frac{1}{2}}\} \leq N\mathcal{E} \\ \qquad \text{rank}\{\boldsymbol{T}\} \leq \theta \end{cases} \quad (17)$$

where we recall $\boldsymbol{T} = \boldsymbol{B}\boldsymbol{\Lambda}\boldsymbol{B}^H$. Since the rank constraint forces $\lambda_{\theta+1} = \ldots = \lambda_N = 0$, (17) is also equivalent to

$$\begin{cases} \max_{\boldsymbol{\Lambda}_\theta \succeq 0,\, \boldsymbol{B}_\theta^H \boldsymbol{B}_\theta = \boldsymbol{I}_\theta} \sum_{j=1}^{\theta} \log_2(1+\sigma_\alpha^2 \lambda_j) \\ \text{s.t.} \quad \text{tr}\{\boldsymbol{B}_\theta \boldsymbol{\Lambda}_\theta \boldsymbol{B}_\theta^H \boldsymbol{M}\} \leq N\mathcal{E} \end{cases} \quad (18)$$

where $\boldsymbol{\Lambda}_\theta = \text{diag}\{\lambda_1,\ldots,\lambda_\theta\}$, $\boldsymbol{B}_\theta = (\boldsymbol{b}_1,\ldots,\boldsymbol{b}_\theta)$ and $\boldsymbol{b}_j$ is the $j$-th column of the unitary matrix $\boldsymbol{B} = (\boldsymbol{b}_1,\ldots,\boldsymbol{b}_N)$. Observe that the choice of $\boldsymbol{B}_\theta$ only affects the feasible region of the problem (18). As a consequence, the optimal $\boldsymbol{B}_\theta$ must ensure the largest feasible set for $\boldsymbol{\Lambda}_\theta$ which rules the value of the objective function. This implies that the first $\theta$ columns of $\boldsymbol{B}$ must be equal to the eigenvectors $\boldsymbol{u}_{M,1},\ldots,\boldsymbol{u}_{M,\theta}$ of $\boldsymbol{M}$ corresponding to the $\theta$ smallest eigenvalues [10, Lemma 9.H.1.h]. The last $N - \theta$ columns of $\boldsymbol{B}$ can be arbitrarily chosen (provided that $\boldsymbol{B}$ is unitary) since they do not come into play in the optimization (18). Therefore, we can assume in the following that $\boldsymbol{B} = \boldsymbol{U}_M$, which in turn implies

$\boldsymbol{U}_A = \boldsymbol{U}_M$ and $\boldsymbol{T} = \boldsymbol{U}_M \boldsymbol{\Lambda}_M^{-\frac{1}{2}} \boldsymbol{P}_A \boldsymbol{\Lambda}_M^{-\frac{1}{2}} \boldsymbol{U}_M^H$ with $\boldsymbol{P}_A = \boldsymbol{\Sigma}_A \boldsymbol{\Sigma}_A^H = \operatorname{diag}\{p_{A,1}, \ldots, p_{A,\theta}, 0 \ldots, 0\}$. Hence, (18) reduces to the following standard water-filling problem [4]

$$\begin{cases} \max_{p_{A,1} \geq 0, \ldots, p_{A,\theta} \geq 0} \sum_{j=1}^{\theta} \left(1 + \frac{\sigma_\alpha^2 p_{A,j}}{\lambda_{M,j}}\right) \\ \text{s.t.} \quad \sum_{j=1}^{\theta} p_{A,j} \leq N\mathcal{E} \end{cases} \quad (19)$$

whose solution is simply given by

$$p_{A,j} = \max\left(0, \mu - \frac{\lambda_{M,j}}{\sigma_\alpha^2}\right), \quad \sum_{j=1}^{\theta} p_{A,j} = N\mathcal{E}. \quad (20)$$

Some remarks are now in order. Under $\operatorname{tr}\{\boldsymbol{A}\boldsymbol{A}^H\} \leq N\mathcal{E}$ and $\operatorname{rank}\{\boldsymbol{A}\} \leq \theta$, the optimal code matrix $\boldsymbol{A} = \boldsymbol{U}_A \boldsymbol{\Sigma}_A \boldsymbol{V}_A^H$ is recovered (up to a right multiplication by an $M \times M$ unitary matrix) by forcing $\boldsymbol{U}_A = \boldsymbol{U}_M$, $[\boldsymbol{\Sigma}_A]_{j,j} = \sqrt{p_{A,j}}$ for $j = 1, \ldots, \theta$ and $p_{A,j}$ given by (20), and $[\boldsymbol{\Sigma}_A]_{j,j} = 0$ for $j = \theta+1, \ldots, \Delta$. This solution (which subsumes the results in [3] for the special cases $\theta = 1$ and $\theta = N \leq M$) has a nice physical interpretation. The optimized code discards the noisiest $N - \theta$ modes defined by $\boldsymbol{M}$ in the signal space. Also, since the remaining $\theta$ orthogonal directions present a different disturbance level, more energy is allocated to more reliable modes according to (20). The number of activated modes depends not only upon the upper rank constraint $\theta$, but also upon the available transmit energy $\mathcal{E}$, the target strength $\sigma_\alpha^2$ and the eigenvalues $\lambda_{M,1} \leq \ldots \leq \lambda_{M,\theta}$ of the disturbance covariance matrix. If $\mathcal{E}\sigma_\alpha^2/\lambda_{M,\theta} \to \infty$, then $p_{A,1} \geq \cdots \geq p_{A,\theta} > 0$ and $\delta = \operatorname{rank}\{\boldsymbol{A}\} = \theta$. Instead, if not enough energy is available for transmission or if the target is weak, additional modes may be switched off. Also, notice that the activated modes have in general different received average signal-to-clutter ratio's given by $\sigma_\alpha^2 p_{A,j}/\lambda_{M,j}$ for $j = 1, \ldots, \delta \leq \theta$ (more on this *infra*).

We emphasize that the situation here is dramatically different from what observed in the previous section under a received average SCR constraint, wherein the transmitter always activates the largest possible number of orthogonal modes permitted by the rank constraint and pumps into them as much energy as to make these paths equivalent at the receiver side.

*1) Analysis and trade-offs:* In Figure 4, we report $1 - P_d$ (probability of miss) for the optimal water-filling solution in (20). We assume $P_{fa} = 10^{-4}$, $M = N = 4$ and $L = 1$; also, we consider an exponentially-shaped disturbance covariance matrix with $[\boldsymbol{M}]_{i,j} = \sigma_d^2 0.7^{|i-j|}$, whose eigenvalues are $\{0.203, 0.318, 0.754, 2.725\}\sigma_d^2$. To elicit the joint effects of the transmit energy constraint and the target strength, the curves are plotted versus the parameter $\gamma = \mathcal{E}\sigma_\alpha^2/\sigma_d^2$.

Notice first that, if $\gamma \to 0$, the optimal code matrix activates only the least interfered mode, *independent* of the rank constraint $\theta$; in this case, the test statistic (6) is a Gamma random variable with shape parameter $L$ and scale parameter

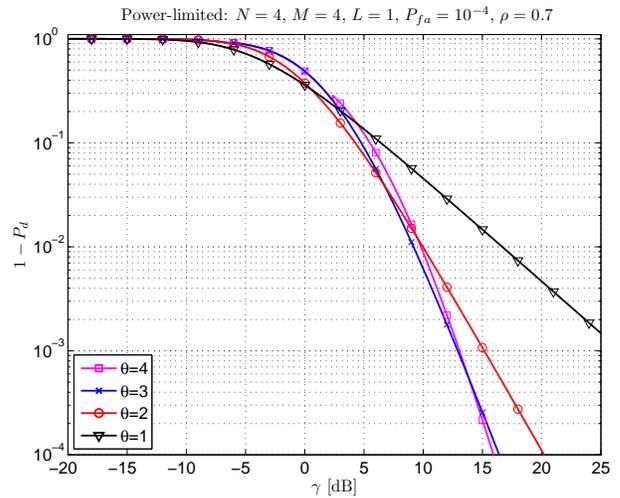

Fig. 4. Probability of miss versus $\gamma$ for the optimal coding in (20).

$1 + \sigma_\alpha^2 N\mathcal{E}/\lambda_{M,1}$ under $H_1$, implying that

$$P_d = \exp\left\{-\frac{\eta}{1 + \sigma_\alpha^2 N\mathcal{E}/\lambda_{M,1}}\right\} \sum_{\ell=0}^{L-1} \frac{1}{\ell!} \left(\frac{\eta}{1 + \sigma_\alpha^2 N\mathcal{E}/\lambda_{M,1}}\right)^\ell.$$

This result confirms the basic intuition that, when the transmit energy is scarce or if the target is weak, we should give up diversity and concetrate all the energy along the least interfered direction in the signal space in order to maximize the received average signal-to-disturbance ratio. A similar result also holds in MIMO point-to-point wireless channels [6].

On the other hand, if $\gamma$ is large enough, relaxing the rank constraint $\theta$ allows generating $\delta \geq 1$ orthogonal modes[3]. In our example, if $-6.1\,\mathrm{dB} < \gamma < 2.3\,\mathrm{dB}$ up to three modes can be activated; instead, full rank coding becomes possible only if $\gamma > 2.3$ dB. For a fixed $\mathcal{E}$, it is clear that increasing the number $\delta$ of modes (i.e., diversity branches) again cames at the price of reducing the energy transmitted (and therefore the average SCR ratio received) along each path. As a consequence, there is no strategy which is uniformly optimal for all values of $\gamma$.

## REFERENCES

[1] E. Fishler, A. Haimovich, R. Blum, L. Cimini, D. Chizhik, and R. Valenzuela, "Spatial diversity in radars - Models and detection performance," *IEEE Trans. Signal Process.*, vol. 54, no. 3, pp. 823–838, Mar. 2006.
[2] A. De Maio and M. Lops, "Design principles of MIMO radar detectors," *IEEE Trans. Aerosp. Electron. Syst.*, vol. 43, no. 3, pp. 886–898, Jul. 2007.
[3] ——, "Space-time coding in MIMO radar," in *MIMO Radar Signal Processing*, Jian-Li and P. Stoica, Eds. New York, NY: John Wiley & Sons, Inc., Apr. 2008 (in press).
[4] T. M. Cover and J. A. Thomas, *Elements of Information Theory*. New York, NY: John Wiley & Sons, Inc., 1991.
[5] M. R. Bell, "Information theory and radar waveform design," *IEEE Trans. Inf. Theory*, vol. 39, no. 5, pp. 1578–1597, Sep. 1993.
[6] D. Tse and P. Viswanath, *Fundamentals of Wireless Communication*. Cambridge University Press, 2005.
[7] H. L. V. Trees, *Detection, Estimation, and Modulation Theory – Part III*. New York, NY: John Wiley & Sons, Inc., 2001.
[8] E. J. Kelly and W. L. Root, "A representation of vector-valued random processes," *Journal of Mathematics and Physics*, vol. 39, no. 3, pp. 211–216, Oct. 1960.

---

[3]In this case, the test statistic (6) is the sum of $\delta$ Gamma random variables with shape parameter $L$ and different scale parameters.

[9] R. A. Horn and C. R. Johnson, *Matrix Analysis*. Cambridge University Press, 1993.
[10] A. W. Marshall and I. Olkin, *Inequalities: Theory of Majorization and Its Applications*. Academic Press, 1979.